\documentclass[twocolumn,twocolappendix,usenatbib,useAMS,fleqn,usenames,dvipsnames, tighten]{aastex63}
\usepackage{amsmath}
\hypersetup{linkcolor=MidnightBlue,citecolor=MidnightBlue,filecolor=cyan,urlcolor=MidnightBlue}
\graphicspath{{./}{plots/}}
\submitjournal{MNRAS}

\usepackage{enumitem, amsmath,amstext, mathtools}

\newcommand{\datastatement}[1]{\begin{small}\section*{Data Availability Statement}\end{small}{\noindent #1}\vspace{5pt}}

\begin{document}

\shorttitle{Cold Gas Subgrid Model}
\shortauthors{Butsky et al.}
\title{Cold Gas Subgrid Model (CGSM): A Two-Fluid Framework for Modeling Unresolved Cold Gas in Galaxy Simulations}
\correspondingauthor{Iryna S. Butsky}
\email{ibutsky@stanford.edu}

\author[0000-0003-1257-5007]{Iryna S. Butsky}\thanks{NASA Hubble Fellow}
\affiliation{Kavli Institute for Particle Astrophysics and Cosmology and Department of Physics, Stanford University, Stanford, CA 94305, USA}
\affiliation{TAPIR, California Institute of Technology, Pasadena, CA 91125, USA}
\affiliation{Astronomy Department, University of Washington, Seattle, WA 98195, USA}
\author[0000-0002-3817-8133]{Cameron B. Hummels}
\affiliation{TAPIR, California Institute of Technology, Pasadena, CA 91125, USA}
\author[0000-0003-3729-1684]{Philip F. Hopkins}
\affiliation{TAPIR, California Institute of Technology, Pasadena, CA 91125, USA}
\author[0000-0001-5510-2803]{Thomas R. Quinn}
\affiliation{Astronomy Department, University of Washington, Seattle, WA 98195, USA}
\author[0000-0002-0355-0134]{Jessica K. Werk}
\affiliation{Astronomy Department, University of Washington, Seattle, WA 98195, USA}

% Abstract of the paper
\begin{abstract}
The cold ($\sim 10^{4}\,{\rm K}$) component of the circumgalactic medium (CGM) accounts for a significant fraction of all galactic baryons. However, using current galaxy-scale simulations to determine the origin and evolution of cold CGM gas poses a significant challenge, since it is computationally infeasible to directly simulate a galactic halo alongside the sub-pc scales that are crucial for understanding the interactions between cold CGM gas and the surrounding ``hot'' medium. In this work, we introduce a new approach: the Cold Gas Subgrid Model (CGSM), which models unresolved cold gas as a second fluid in addition to the standard ``normal'' gas fluid. The CGSM tracks the total mass density and bulk momentum of unresolved cold gas, deriving the properties of its unresolved cloudlets from the resolved gas phase. The interactions between the subgrid cold fluid and the resolved fluid are modeled by prescriptions from high-resolution simulations of ``cloud crushing'' and thermal instability. Through a series of idealized tests, we demonstrate the CGSM's ability to overcome the resolution limitations of traditional hydrodynamics simulations, successfully capturing the correct cold gas mass, its spatial distribution, and the timescales for cloud destruction and growth. We discuss the implications of using this model in cosmological simulations to more accurately represent the microphysics that govern the galactic baryon cycle.
\end{abstract}


\section{Introduction}

Galactic evolution is driven by the exchange of baryons between the stellar disk and its surrounding halo. The interstellar medium (ISM) requires a consistent inflow of gas to form stars, which then expel chemically-enriched gas back into the circumgalactic medium (CGM) at the end of their lifespans. In this way, the composition of the CGM provides a detailed record of a galaxy's past and the physical processes that govern its evolution \citep{Tumlinson:2017, FaucherGiguere:2023}. In particular, cold ($\sim 10^4\,{\rm K}$) CGM gas is a crucial component fueling ongoing star formation.

Cold gas is ubiquitous in the halos around dwarf, $L_*$, and massive galaxies across cosmic time \citep[e.g.][]{Thom:2012, Werk:2013, Cantalupo:2014, Borisova:2016, Mishra:2022, Qu:2022}. This cold gas plays an integral role in the galactic baryon cycle and is frequently observed both, in galactic outflows \citep[e.g.][]{Heckman:2000, Nielsen:2015, Veilleux:2020, Burchett:2021} and in accretion streams into the galaxy \citep{Sancisi:2008, Putman:2012, Richtler:2018, Zhu:2021, Kamphuis:2022}. Recent estimates indicate that cold CGM gas constitutes at least 25\% of the baryonic content in $L_{\ast}$ galaxies \citep{Werk:2014}, and in dwarf galaxies, it can account for as much as 90\% of galactic baryons \citep{Zhang:2020}. Consequently, to fully understand the galactic baryon cycle and galaxy evolution, theoretical models must not only reproduce the observed properties of cold CGM gas but also provide robust predictions about the physical processes influencing cold gas formation, destruction, and its interactions with the hot phase.

Recent theoretical work and high-resolution idealized simulations have significantly advanced our understanding of the formation and evolution of cold gas across various contexts. For instance, it is now established that when radiative cooling is sufficiently efficient, cold gas can condense from thermally-unstable hot gas \citep{Field:1965, McCourt:2012, Sharma:2012a, Girichidis:2021} and subsequently precipitate onto the host galaxy \citep{Voit:2015}. This thermal instability has been explored in diverse physical contexts including thermal conduction \citep{Sharma:2010, Wagh:2014}, magnetic fields \citep{Ji:2018}, turbulence \citep{Voit:2018}, and cosmic rays \citep{Butsky:2020, Tsung:2023}. Additionally, high-resolution ``cloud-crushing'' simulations have elucidated the conditions that lead to either the destruction or mass accretion of cold gas clouds \citep{Armillotta:2017, Gronke:2018, Li:2020, Gronke:2020b, Sparre:2020, Kanjilal:2021, Farber:2022, Fielding:2022, Abruzzo:2023}, as influenced by the thin, radiatively cooling boundary layer interface between the hot and cold gas phases \citep{Fielding:2020, Tan:2021, Chen:2023}. Recent theoretical advances have begun to frame our understanding of what determines the size of cold gas clouds and under what circumstances these clouds form a mist of cloudlets or coalesce into larger clouds \citep{Gronke:2020b, Gronke:2023}. However, without a cosmological context \citep[e.g.][]{Saeedzadeh:2023}, determining the relative impact of each of these processes on the observed structures of cold CGM gas is exceedingly difficult.

Studying cold CGM gas with galaxy-scale simulations presents significant challenges due to the vast range of relevant physical scales involved. Theoretically, the characteristic scale of cold CGM clouds is estimated to be around $0.1-10$ pc \citep{McCourt:2018, Gronke:2018, Liang:2020}. While it is difficult to observationally determine the \textit{minimum} size of these clouds, there is evidence suggesting the existence of cold CGM clouds smaller than $\sim 10-100$ pc \citep{Hennawi:2015, Stern:2016, Rubin:2018, Bish:2019, Rudie:2019, Werk:2019, Zahedy:2019, HWChen:2023}. The exact resolution required to accurately simulate cold CGM gas in a galactic context is still under debate. However, even simulations that employ innovative algorithms to maximize CGM resolution (reaching down to about $100$ pc in the inner CGM) have not achieved convergence across all relevant cold gas properties \citep{Hummels:2019, Peeples:2019, Suresh:2019, vandeVoort:2019, Nelson:2020, Ramesh:2024}. 

If it is necessary to resolve scales significantly smaller than 100 pc, we may be decades away from the computational resources required to explicitly simulate cold CGM gas physics. For example, resolving the halo of an $L_{\ast}$ galaxy out to $200\, {\rm kpc}$ with parsec-scale spatial resolution would require roughly $10$ million times more voxels than current state-of-the-art cosmological simulations. However, the current alternative of studying CGM gas using simulations that severely underresolve its structure substantially restricts the conclusions we can draw regarding the origins of cold gas and its impact on galaxy evolution. 

Instead of waiting for technological advances that enable direct resolution of the CGM at sub-pc scales, we will show that resolution issues can be effectively addressed using a physically-motivated subgrid model for cold CGM gas. The use of subgrid models has a strong precedent in galaxy simulations, particularly for processes that cannot be directly resolved, such as star formation and supernova feedback \citep[e.g.][]{Cen:1992, Abadi:2003, Oppenheimer:2006, Stinson:2006, Hopkins:2012, Keller:2014}, ISM phase structure \citep[e.g.][]{Springel:2003a}, and metal diffusion \citep[e.g.][]{Shen:2010, Escala:2017}. More recently, several studies have developed subgrid prescriptions to model unresolved cold gas in multiphase supernova-driven winds \citep{Huang:2020, Huang:2022, Smith:2024}. 

In this work, we present a novel framework for explicitly modeling unresolved cold CGM gas in hydrodynamics simulations, conceptually similar to how dust is modeled in protoplanetary disks \citep[e.g.][]{Laibe:2012a} or the more general multifluid cosmological hydrodynamics framework \citep{Weinberger:2023}. Our Cold Gas Subgrid Model (CGSM), treats this unresolved cold gas as a second fluid, tracking its total mass density and bulk momentum as collections of unresolved cold cloudlets. The properties of these cloudlets are determined by the properties of the resolved (``hot'') gas phase, which we assume to be in thermal pressure equilibrium with the unresolved cold gas. The interaction terms between the standard and subgrid fluids are informed by the latest models for cold gas formation and destruction, as indicated by the idealized simulations mentioned above. As we will demonstrate, this subgrid approach effectively replicates the qualitative behavior of cold gas in situations where traditional hydrodynamics simulations struggle due to inadequate resolution.

The remainder of this paper is organized as follows: In \autoref{sec:model}, we detail the physical basis of the CGSM and introduce the set of modified conservation equations governing the evolution of the two fluids. In \autoref{sec:application}, we use a series of idealized simulations to show that, in contrast to traditional hydrodynamics, our two-fluid model successfully captures the correct qualitative behavior of cold CGM gas, even in cases of limit of very low resolution. The implications of the CGSM for cosmological simulations are discussed in \autoref{sec:discussion}, and we summarize our results in \autoref{sec:summary}. Finally, in \autoref{sec:appendix}, we provide tests of the CGSM implementation in the {\sc enzo} astrophysical simulation code \citep{Bryan:2014, Enzo:2019}. 

\section{Cold Gas Subgrid Model Two-Fluid Description}\label{sec:model}
\begin{figure}
\includegraphics[width=0.45\textwidth]{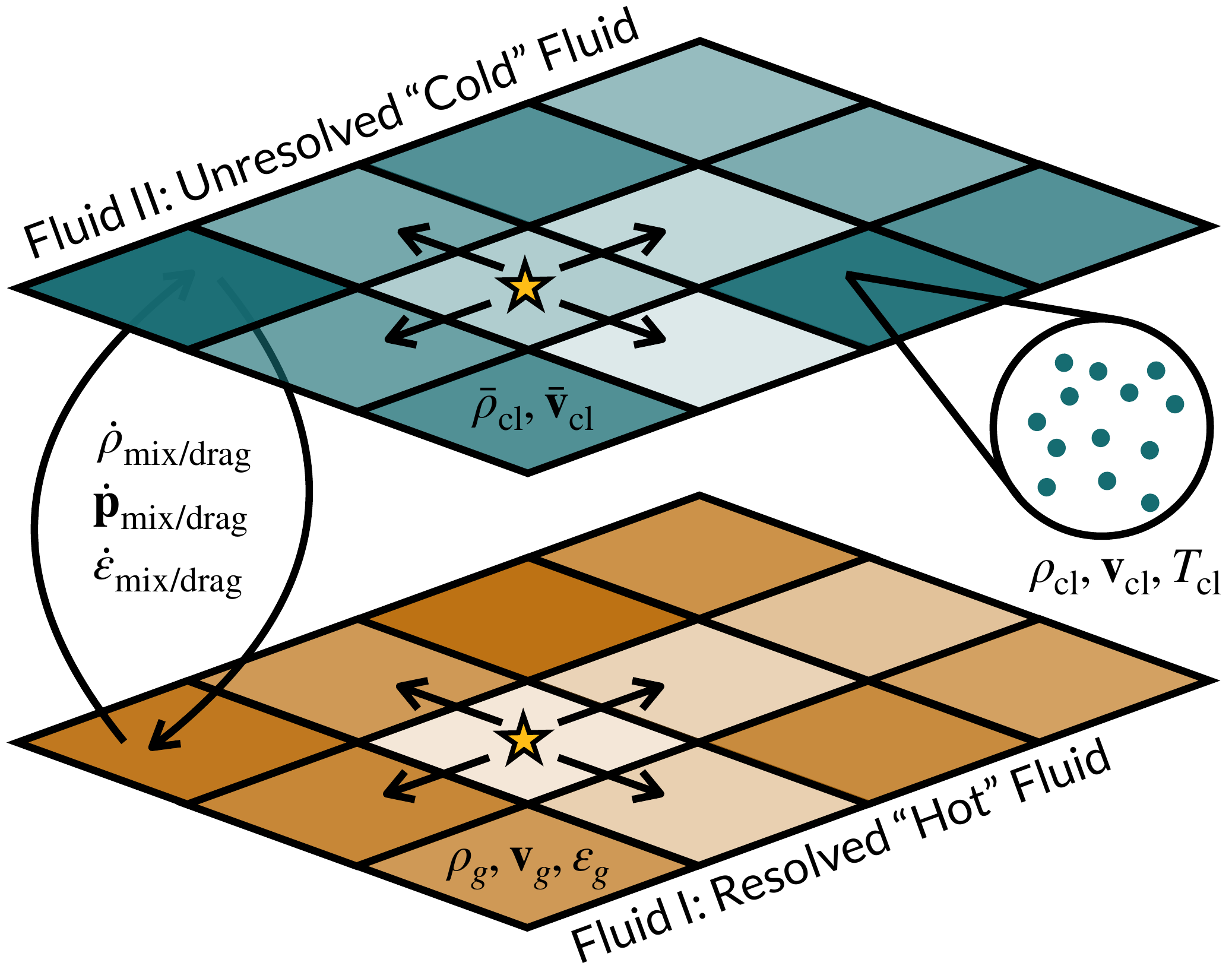}
\caption{This schematic illustrates the CGSM.The resolved ``hot'' fluid (bottom) functions as a typical hydrodynamic fluid, with each cell tracking several key properties (gas density $\rho_g$, velocity ${\bf v}_g$, and energy density $\varepsilon_g$) that evolve over time following the conservation equations outlined in equations \ref{eqn:mass_new} - \ref{eqn:energy_new}. The second, unresolved ``cold'' fluid (top) is spatially co-existent with the resolved fluid. This second fluid explicitly tracks the total mass density $\bar{\rho}_{\rm cl}$ and bulk velocity (${\bf \bar{v}_{\rm cl}}$) of the unresolved cold cloudlets. Interactions between these two fluids, including the exchange of mass, momentum, and energy occur via the mixing and drag terms described in \autoref{sec:model}. The yellow stars symbolize injected sources of gas, such as those resulting from stellar feedback.} 
\label{fig:diagram}
\end{figure}

In this section, we detail the physical rationale and numerical implementation of our Cold Gas Subgrid Model, as illustrated in \autoref{fig:diagram}. Briefly stated, the resolved (hot) gas phase is treated as a conventional hydrodynamics fluid, adhering to the equations outlined in \autoref{sec:traditional} and modified in \autoref{sec:cgsm_equations}. Meanwhile, the unresolved (cold) gas phase is represented as a second fluid, co-spatial with the hot fluid, following the set of modified governing equations described in \autoref{sec:cgsm_equations}. Given that the structure of the cold phase is not directly resolvable, we focus on evolving the total mass density and bulk momentum of this phase. We make informed assumptions about the physical characteristics of the unresolved cold gas based on the properties of its surrounding medium. This approach is qualitatively similar to modeling dust in protoplanetary disks as described in \citealt{Laibe:2012a}, as well as the compressible multifluid hydrodynamics described in \citealt{Weinberger:2023}.

\subsection{Traditional ``Resolved'' Fluid Equations} \label{sec:traditional}
The evolution of the resolved gas fluid is described by the conventional hydrodynamics conservation equations for mass, momentum, and energy: 

\begin{equation}\label{eqn:mass}
\frac{\partial \rho_g}{\partial t} + \nabla \cdot (\rho_g {\bf v}_g) = 0,
\end{equation}

\begin{equation}\label{eqn:momentum}
\frac{\partial(\rho_g {\bf v}_g)}{\partial t} + \nabla 
	\cdot (\rho_g{\bf v}_g {\bf v}_g^T + {\bf I}P_g) = -\rho_g \nabla {\bf \Phi},
\end{equation}

\begin{equation}\label{eqn:energy}
\frac{\partial \varepsilon_g}{\partial t} + \nabla\cdot ({\bf v}_g \varepsilon_g)
= - P_g\nabla\cdot{\bf v}_g + S_g.
\end{equation}
Here, $\Phi$ represents the gravitational potential, while $\rho_g$ and ${\bf v}_g$ denote the gas density and velocity, respectively. {\bf I} is the identity matrix and the internal gas pressure, $P_g$, is related to the internal thermal energy density $\varepsilon_g = (\gamma_g - 1)P_g$, where $\gamma_g = 5/3$. $S_g$ symbolizes thermal energy sources and sinks, such as energy inputs from supernova feedback events or energy losses due to radiative cooling. Throughout, the subscript ``$g$'' indicates the properties of the resolved gas. The standard equations above are provided for context, leading to the modified conservation equations introduced in equations \ref{eqn:mass_new} - \ref{eqn:energy_new}. 

\subsection{Unresolved Cold Gas Fluid}
In addition to evolving a ``traditional'' resolved fluid, the CGSM explicitly evolves a second fluid of cold gas, comprised of numerous unresolved cloudlets, under the following assumptions:

First, we assume that the size of the unresolved cold gas cloudlets is significantly smaller than the resolution element size, $r_{\rm cl} \ll \Delta x_{\rm cell}$, an assumption easily met in the simulated CGM with cell widths $\Delta x \gtrsim 1$ kpc. With this disparity in scale, we make the simplifying assumption that the cold gas clouds are in thermal pressure equilibrium with the resolved hot phase and ignore any pressure gradients within the cold cloud interiors. This allows us to treat the unresolved gas as a pressure-free fluid. Given that the majority of thermal pressure originates from the hot gas phase, we can deduce the system's thermal pressure from the properties of the resolved fluid. 

Second, we assume that the unresolved cold fluid is at a constant temperature $T_{\rm cl}$. Unless otherwise noted, we default to $T_{\rm cl} = 10^4\, {\rm K}$. 
Assuming the two fluids are in thermal pressure equilibrium, the physical density of the unresolved cloudlets is given by:
\begin{equation}\label{eqn:cloud_density}
  \rho_{\rm cl} = \frac{\rho_g T_{g}}{ T_{\rm cl}} = \frac{\mu m_p P_g}{k_B T_{\rm cl}}, 
\end{equation}
where $T_g$ is the temperature of the resolved gas phase, $\mu$ is the mean molecular weight, $m_p$ is the proton mass, and $k_B$ is the Boltzmann constant. The subscript ``${\rm cl}$'' is used throughout to refer to the properties of the cold gas fluid.

Third, we assume that the characteristic size of cold cloudlets is $r_{\rm cl} \approx \mathrm{min}(c_s t_{\rm cool})$, motivated by the mist model of cold CGM gas \citep{McCourt:2018}.
The gas sound speed and cooling time are described by 
\begin{equation}
    c_s = \sqrt{(\gamma P_g / \rho_g)},
\end{equation}
and 
\begin{equation}
t_{\rm cool} = P_g / (\gamma - 1) \mathcal{L},
\end{equation}
for some radiative gas cooling rate $\mathcal{L}$. For a wide range of CGM pressures, the minimum product of the sound speed and cooling time occurs at a temperature of $T_{\rm cool, peak} \approx 10^{4.3}\,{\rm K}$ \citep{Liang:2020}. Therefore, we solve for the cold cloud radius at $T = T_{\rm cool, peak}$, 
\begin{equation}\label{eqn:cloud_radius}
r_{\rm cl} = (c_s t_{\rm cool}) |_{T_{\rm cool, peak}}.
\end{equation}
It is important to note that both the gas sound speed and cooling time in the above equation are calculated at $T = T_{\rm cool, peak}$. Assuming thermal pressure equilibrium, we solve for the gas density at the peak-cooling temperature, $\rho_{\rm cool, peak} / \rho_g = T_g / T_{\rm cool, peak}$. In the simulations below, we assume $T_{\rm cool, peak} = 10^{4.3}\, {\rm K}$.

Combining equations \ref{eqn:cloud_density} and \ref{eqn:cloud_radius}, we estimate the mass of a single cold gas cloudlet as
\begin{equation}\label{eqn:cloud_mass}
    m_{\rm cl} = \frac{4}{3} \pi r_{\rm cl}^3 \rho_{\rm cl}. 
\end{equation}
It then follows that the number of cold cloudlets in a given cell is given by, 
\begin{equation}\label{eqn:cloud_number}
N_{\rm cl} = M_{\rm cl} / m_{\rm cl}, 
\end{equation}
where $M_{\rm cl}$ is the total mass of unresolved cold gas in a cell. 

\subsection{Conservation Equations for the CGSM}\label{sec:cgsm_equations}
We explicitly evolve the total mass density ($\bar{\rho}_{\rm cl}$) and bulk velocity (${\bf \bar{v}_{\rm cl}}$) of the unresolved cold fluid in each cell. Since the temperature of the cold gas cloudlets is assumed to be constant, we do not explicitly evolve the internal energy of the cold gas fluid. In the following sections, we introduce the new conservation equations for the unresolved fluid and outline the necessary modifications for the resolved fluid.

\subsubsection{Conservation of Mass}
\begin{equation}\label{eqn:mass_new}
\frac{\partial \rho_g}{\partial t} + \nabla \cdot (\rho_g {\bf v}_g) = -\dot{\rho}_{\rm mix},
\end{equation}

\begin{equation}\label{eqn:mass_cold_new}
\frac{\partial \bar{\rho}_{\rm cl}}{\partial t} + \nabla \cdot (\bar{\rho}_{\rm cl} {\bf \bar{v}_{\rm cl}}) = \dot{\rho}_{\rm mix}.
\end{equation}
Equations \ref{eqn:mass_new} and \ref{eqn:mass_cold_new} function similarly to the traditional mass conservation equation (equation \ref{eqn:mass}), but they include an additional source term, $\dot{\rho}_{\rm mix}$ to account for the mass transfer between the two fluids. This transfer occurs during the formation of cold clouds, their mass accretion, and their mass loss due to hydrodynamic instabilities and conduction. We provide specific prescriptions for these processes in sections \ref{sec:model_TI} and \ref{sec:cloud_crushing}. In the equations above, the variable $\bar{\rho}_{\rm cl} = {\rm M}_{\rm cl, cell} / {\rm V}_{\rm cell}$ represents the total mass of cold clouds in a single cell divided by the cell's volume. It is important to note that this does not represent the physical density of the unresolved cold clouds, $\rho_{\rm cl}$ (equation \ref{eqn:cloud_density}). Similarly, ${\bf \bar{v}_{\rm cl}}$ denotes the aggregate velocity of the unresolved cold fluid in a given cell rather than the physical velocity of any individual cloudlet. 

\subsubsection{Conservation of Momentum}

\begin{equation}\label{eqn:momentum_new}
\begin{split}
\frac{\partial(\rho_g {\bf v}_g)}{\partial t} + \nabla 
	\cdot (\rho_g{\bf v}_g{\bf v}_g^T + {\bf I}P_g) = -\rho_g \nabla {\bf \Phi} -\dot{\bf p}_{\rm mix} - \dot{\bf p}_{\rm drag},
\end{split}
\end{equation}

\begin{equation}\label{eqn:momentum_cold_new}
\begin{split}
\frac{\partial(\bar{\rho}_{\rm cl} {\bf \bar{v}_{\rm cl}})}{\partial t} + \nabla 
	\cdot (\bar{\rho}_{\rm cl}{\bf \bar{v}_{\rm cl}}{\bf \bar{v}_{\rm cl}}^T) = 
 -\bar{\rho}_{\rm cl} \nabla {\bf \Phi} + \dot{\bf p}_{\rm mix} + \dot{\bf p}_{\rm drag}.
\end{split}
\end{equation}
As with the mass conservation equations, the revised momentum conservation equations include new source terms to account for momentum transfer between the cold and hot fluids. These transfers occur due to the formation, growth, and destruction of cold clouds (${\bf \dot{p}}_{\rm mix}$), as well as from the drag force resulting from the relative motion of the two fluids (${\bf \dot{p}}_{\rm drag}$).

The momentum transfer due to mixing of the two fluids is dependent on the mass transfer, $\dot{\rho}_{\rm mix}$, and is expressed as follows:
\begin{equation}
        \dot{\bf p}_{\rm mix} = 
\begin{cases}
    \dot{\rho}_{\rm mix}{\bf v}_g,& \dot{\rho}_{\rm mix} > 0\\
    \dot{\rho}_{\rm mix}{\bf \bar{v}_{\rm cl}},              & \dot{\rho}_{\rm mix} < 0
\end{cases}.
\end{equation}
We note that $\dot{\bf p}_{\rm mix}$ only accounts for the \textit{net} transfer of mass, either to the cold fluid of from the cold fluid. In reality, in the ensemble of unresolved cold clouds within a single cell, some clouds would be losing mass while others would be accreting mass. This could lead an underestimate of the true momentum transfer, for example, in the limit where $\dot{\rho}_{\rm mix} = 0$. We leave exploring these effects to future work.

We model the drag force as a function of the relative velocity between the two fluids, ${\bf v}_{\rm rel} = {\bf v}_g - {\bf \bar{v}_{\rm cl}}$, 
\begin{equation}
    {\bf \dot{p}}_{\rm drag} = K_{\rm drag} {\bf v}_{\rm rel},
\end{equation}
where $K_{\rm drag}$ is the drag coefficient given by: 
\begin{equation}
K_{\rm drag} = \pi r_{\rm cl}^2 \rho_g \frac{\bar{\rho}_{\rm cl}}{m_{\rm cl}} |{\rm v}_{\rm rel}|, 
\end{equation}
as described in equation 41 of \citealt{Laibe:2012b}.

\subsubsection{Conservation of Energy}

\begin{equation}\label{eqn:energy_new}
\frac{\partial \varepsilon_g}{\partial t} + \nabla\cdot ({\bf v}_g \varepsilon_g) = - P_g\nabla\cdot{\bf v}_g + S_g - \dot{\varepsilon}_{\rm mix} - \dot{\varepsilon}_{\rm drag}.
\end{equation} 
Since the cold gas fluid has a fixed temperature, we do not explicitly track its energy field.However, the energy conservation equation for the traditional fluid is modified to include two additional source terms, $\dot{\varepsilon}_{\rm mix}$ and $\dot{\varepsilon}_{\rm drag}$. The energy lost from the resolved fluid due to mixing with the cold phase is calculated as
\begin{equation}
\dot{\varepsilon}_{\rm mix} = \dot{\rho}_{\rm mix}(\varepsilon_g / \rho_g).
\end{equation}

In addition, we consider energy loss due to frictional heating caused by drag between the two fluids, given by
\begin{equation}
\dot{\varepsilon}_{\rm drag} = K_{\rm drag} {\bf v}_{\rm rel}^2.
\end{equation} 

We provide tests of the implementation of the modified conservation equations in the {\sc enzo} simulation code in \autoref{sec:appendix}.

\subsection{Cold Cloud Formation due to Thermal Instability}\label{sec:model_TI}
A primary mechanism for the spontaneous formation of cold gas from hot gas is thermal instability \citep{Field:1965}. While the cooling times of hot ($T\sim 10^6\, {\rm K}$) CGM gas are generally long, fluctuations in gas density and temperature can trigger a runaway cooling effect. This process results in cold gas condensing from the hot phase \citep[e.g.][]{McCourt:2012, Sharma:2012a, Voit:2015}. 
In the CGSM, we model this phenomenon by first identifying cells likely to experience thermal instability. Then, we convert the energy that would have been lost due to radiative cooling in the resolved fluid to generate mass in the unresolved cold fluid.

At each timestep, we evaluate the cells of the resolved fluid for thermal instability using the following criteria:
\begin{enumerate}
\item The gas temperature falls within the thermally unstable range, $10^4\, {\rm K} \le T \le 10^6\, {\rm K}$. 
\item The size of the cell exceeds the size necessary to resolve a single cold cloud, $\Delta x_{\rm cell} > r_{\rm cl}$
\item The net energy change due to radiative cooling in the cell is negative, meaning it would cool in the absence of the two-fluid model.
\end{enumerate}

If the hot gas cell is not thermally unstable, it is allowed to undergo radiative cooling (or heating) as usual. However if the gas in a cell is determined to be thermally unstable, we transfer mass from the hot gas to the cold subgrid fluid according to the following equation:
\begin{equation}
\dot{\rho}_{\rm mix, TI} = \Delta \rho_{\rm mix, TI} / \Delta t,
\end{equation}
where $\Delta t$ represents the simulation timestep. Assuming conservation of energy, the quantity of mass transferred depends on the amount of energy the hot gas cell is expected to lose due to radiative cooling, $\Delta E_{\rm cool}$: 

\begin{equation}
\begin{split}
\rho_g e_g + \bar{\rho}_{\rm cl}e_{\rm cl} - |\Delta E_{\rm cool}| = \\ (\rho_g - \Delta \rho_{\rm mix, TI})e_g + (\bar{\rho}_{\rm cl} + \Delta \rho_{\rm mix, TI})e_{\rm cl},
\end{split}
\end{equation}

\begin{equation}
   \Delta \rho_{\rm mix, TI} =  \frac{|\Delta E_{\rm cool}|}{e_g - e_{\rm cl}}.
\end{equation}
In the equations above, $e_{\rm hot},\, e_{\rm cold}$ are the specific energies (energy per unit mass) of the hot and cold gas. Note that $\Delta \rho_{\rm mix, TI}$ is always positive because, by definition, $e_g > e_{\rm cl}$. Therefore, in the case of thermal instability, mass is only ever transferred from the hot fluid to the cold. 

We update each cell, $i$, at time $t$, as follows:
\begin{equation}
\begin{split}
    \rho_i^{t+1} = \rho_i^t - (\Delta \rho_{\rm mix, TI})_i,\\
    \bar{\rho}_{{\rm cl}, i}^{\,t+1} = \bar{\rho}_{{\rm cl},i}^{\,\,t} + (\Delta \rho_{\rm mix, TI})_i.
    \end{split}
\end{equation}

\subsection{Cloud Crushing} \label{sec:cloud_crushing}
A cold cloudlet moving relative to a hot background medium may either lose or accrete mass, depending on the physical conditions of the two gas phases and their relative velocities \citep[e.g.][]{Klein:1994, Marinacci:2010, Armillotta:2016, Gronke:2018, BandaBarragan:2020, Li:2020, Sparre:2020, Abruzzo:2022, Kanjilal:2021}.

In our model, we incorporate the mass transfer prescription as derived in equation 36 of \citealt{Fielding:2022}, which effectively captures various regimes of cold gas growth and destruction observed in high-resolution simulations,
\begin{equation}\label{eqn:cloud_crush}
\dot{\rho}_{\rm mix, cc} = 0.6 \bigg( \frac{\bar{\rho}_{\rm cl} {\bf v}_{\rm rel}}{\chi^{1/2} r_{\rm cl}}\bigg) (\xi^{\alpha} - 1).
\end{equation}
In the equation above, $\chi = \rho_{\rm cl} / \rho_g$ represents the density contrast between the cold cloudlets and the hot phase, and $\xi = r_{\rm cl} / ({\bf v}_{\rm turb} t_{\rm cool, peak})$ determines whether the cold gas is accreting or losing mass, with $\alpha = 1/4$ if $\xi \geq 1$ and $\alpha = 1/2$ otherwise. We assume that the turbulent velocity, ${\bf v}_{\rm turb} = 0.1 {\bf v}_{\rm rel}$, to be a constant fraction of the relative velocity between the two fluids. 

In the limit of no cooling, \autoref{eqn:cloud_crush}, implies a cloud-crushing time, 
\begin{equation}\label{eqn:cloud_crush_time}
    t_{\rm cc} = 1.67\,\bigg(\frac{r_{\rm cl}}{{\bf v}_{\rm rel}}\bigg) \bigg(\frac{\rho_{\rm cl}}{\rho_g}\bigg)^{1/2} .
\end{equation}

\subsection{Time Stepping}
Following the approach in \citep{Laibe:2012a}, we incorporate a time step criterion that is a function of the drag coefficient, $K_{\rm drag}$. The time step requirement is given by
\begin{equation}
\Delta t_{\rm drag} = \bigg(\frac{\rho_g \bar{\rho}_{\rm cl}}{K_{\rm drag} (\rho_g + \bar{\rho}_{\rm cl})}\bigg).
\end{equation}

Next, we consider the time constraints imposed by the mass and momentum transfer due to thermal instability and cloud-crushing processes in the CGSM.
\begin{equation}
    \Delta t_{{\rm mix, }\rho} = {\rm min}\bigg(\frac{\rho_g}{\dot{\rho}_{\rm mix}}, \frac{\bar{\rho}_{\rm cl}}{\dot{\rho}_{\rm mix}}\bigg),
\end{equation}
\begin{equation}
    \Delta t_{{\rm mix,v}} = {\rm min}\bigg(\frac{{\bf p}_g}{\dot{\bf p}_{\rm mix}}, \frac{\bar{\bf p}_{\rm cl}}{\dot{\bf p}_{\rm mix}}\bigg).
\end{equation}

The final time step constraint for the CGSM is the minimum of the above constraints, 
\begin{equation}
\Delta t = \epsilon_c {\rm min}(\Delta t_{\rm drag}, \Delta t_{{\rm mix, }\rho}, \Delta t_{\rm mix, v}).
\end{equation}
In the simulations below, we choose the constant prefactor $\epsilon_c = 0.2$.

\section{Physical Applications of the CGSM}\label{sec:application}
Next, we use a series of idealized simulations to underscore the fundamental qualitative differences between traditional hydrodynamics and CGSM simulations. Specifically, we focus on examples of the formation, spatial distribution, destruction, and accretion of cold gas. As we will detail below, a recurring theme is that in the low-resolution limit typical of the CGM in galaxy simulations, traditional hydrodynamics significantly diverges from the ``true'' behavior of cold gas. In such instances, the CGSM is able to reproduce the essential qualitative behavior of cold gas more accurately.

\begin{figure}
\includegraphics[width=0.47\textwidth]{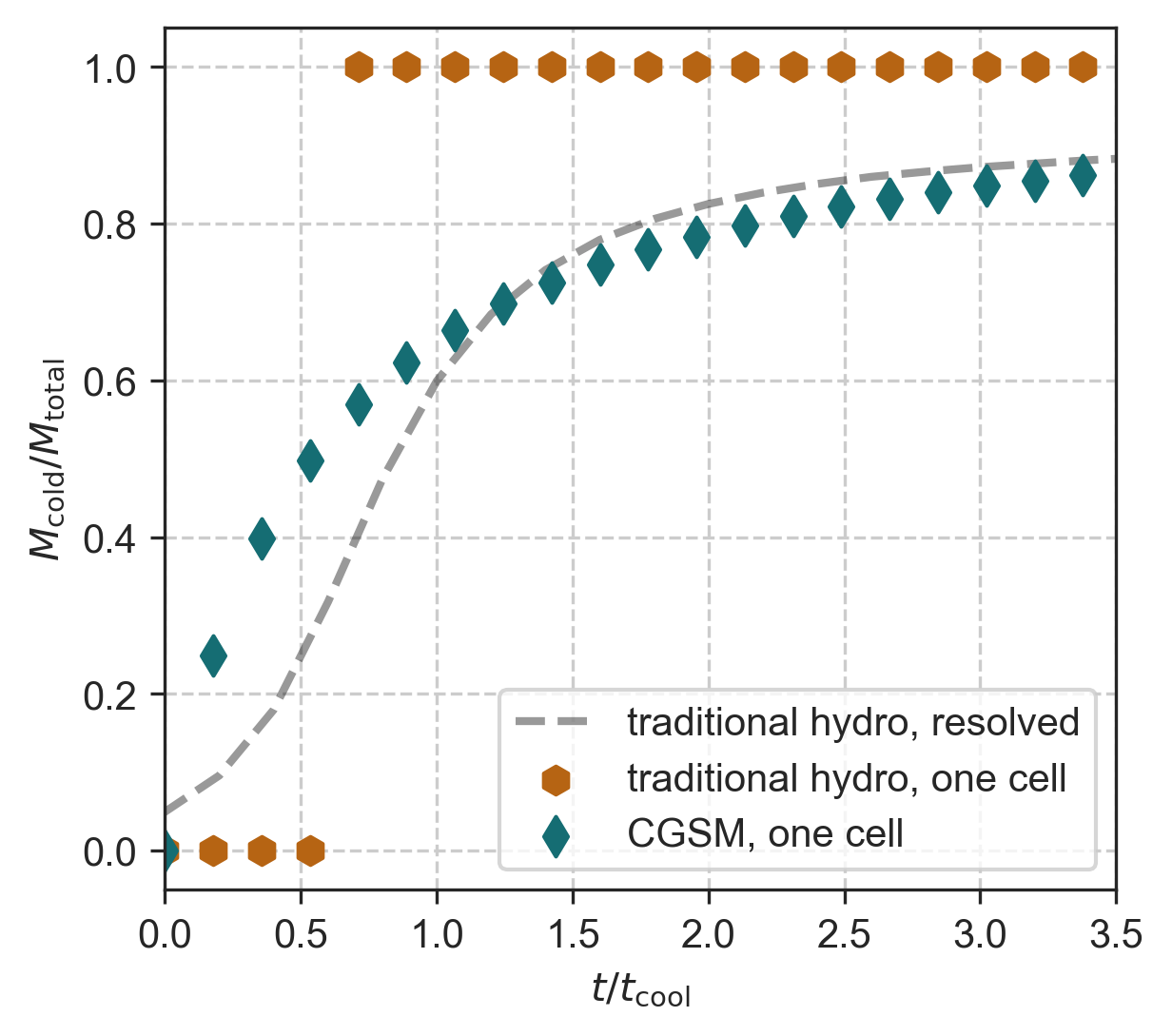}
\caption{The cold mass growth rate in a single cell using traditional hydrodynamics (orange) and the CGSM (blue). In both simulations, the initial condition models a cell of uniform temperature and density undergoing radiative cooling. In the traditional approach, the transition from ``hot'' to ``cold'' gas in the cell is abrupt, resembling a step function. Conversely, the CGSM facilitates a more gradual transfer of mass between the two gas phases within the same cell. For comparison, the dashed line represents the cold gas growth rate in a fully resolved simulation of a thermally unstable patch. Notably, the CGSM captures the spatial coexistence of cold and hot gas phases within a single cell.} 
\label{fig:onezone} 
\end{figure}

\begin{figure*}
\includegraphics[width=\textwidth]{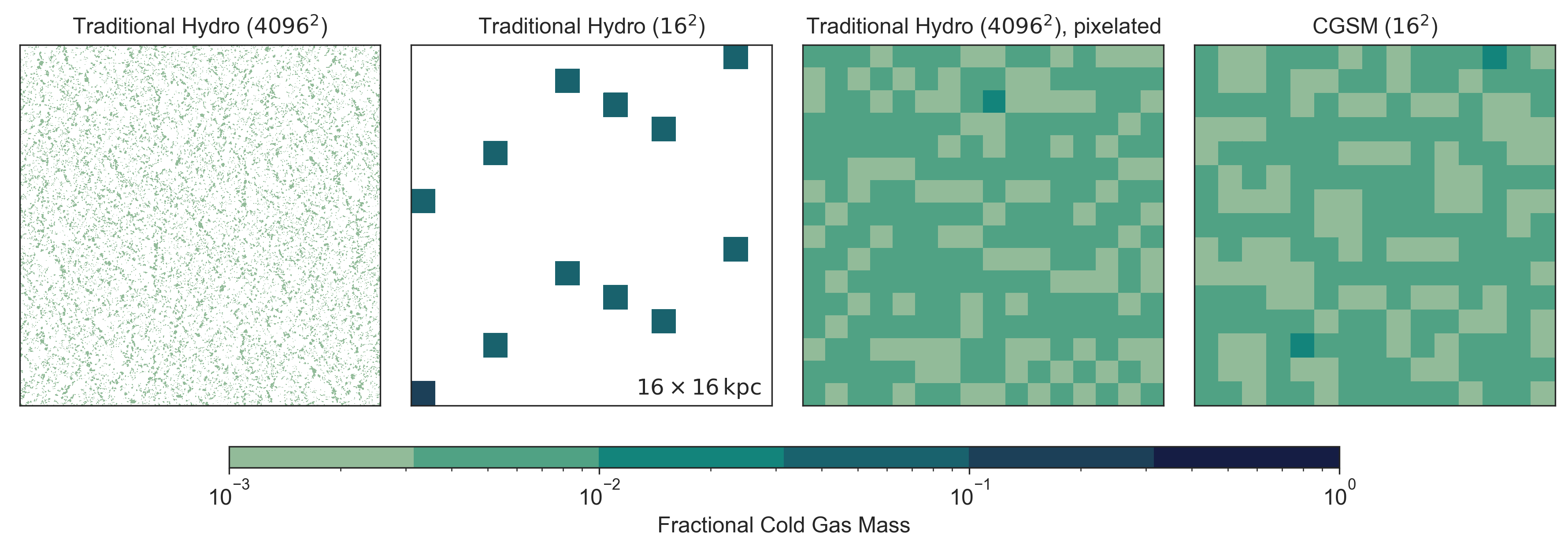}
\caption{The distribution of cold gas formed via thermal instability in idealized 2D simulations of a $16\times16$ kpc patch of CGM gas. Progressing from left to right, the panels show the cold gas mass fraction ($M_{\rm cl} / M_{\rm cl, total}$) in: (1) a high-resolution traditional hydrodynamics simulation with $4096^2$ resolution, (2) a low-resolution traditional hydrodynamics simulation with $16^2$ resolution, (3) the high-resolution simulation from the first panel but binned to a $16^2$ resolution, and (4) a CGSM simulation with $16^2$ resolution. In the high-resolution simulation, cold gas condenses into cloudlets of approximately $r_{\rm cl} \approx 10$ pc in size. However, in the low-resolution traditional hydrodynamics simulation -- where the resolution is $\Delta x = 1$ kpc, akin to typical resolutions in the CGM of galaxy-scale simulations -- the cold gas appears as single-cell clouds with artificially inflated sizes due to cell resolution.This results in a cold gas mass distribution on $\sim$kpc scales that qualitatively differs from what is predicted by the binned high-resolution simulation in the third panel. In contrast, the CGSM simulation successfully captures the correct qualitative distribution of cold gas mass on kpc scales, even without directly resolving the cold cloudlets. } 
\label{fig:ti_comparison} 
\end{figure*}

\begin{figure}
\includegraphics[width=0.49\textwidth]{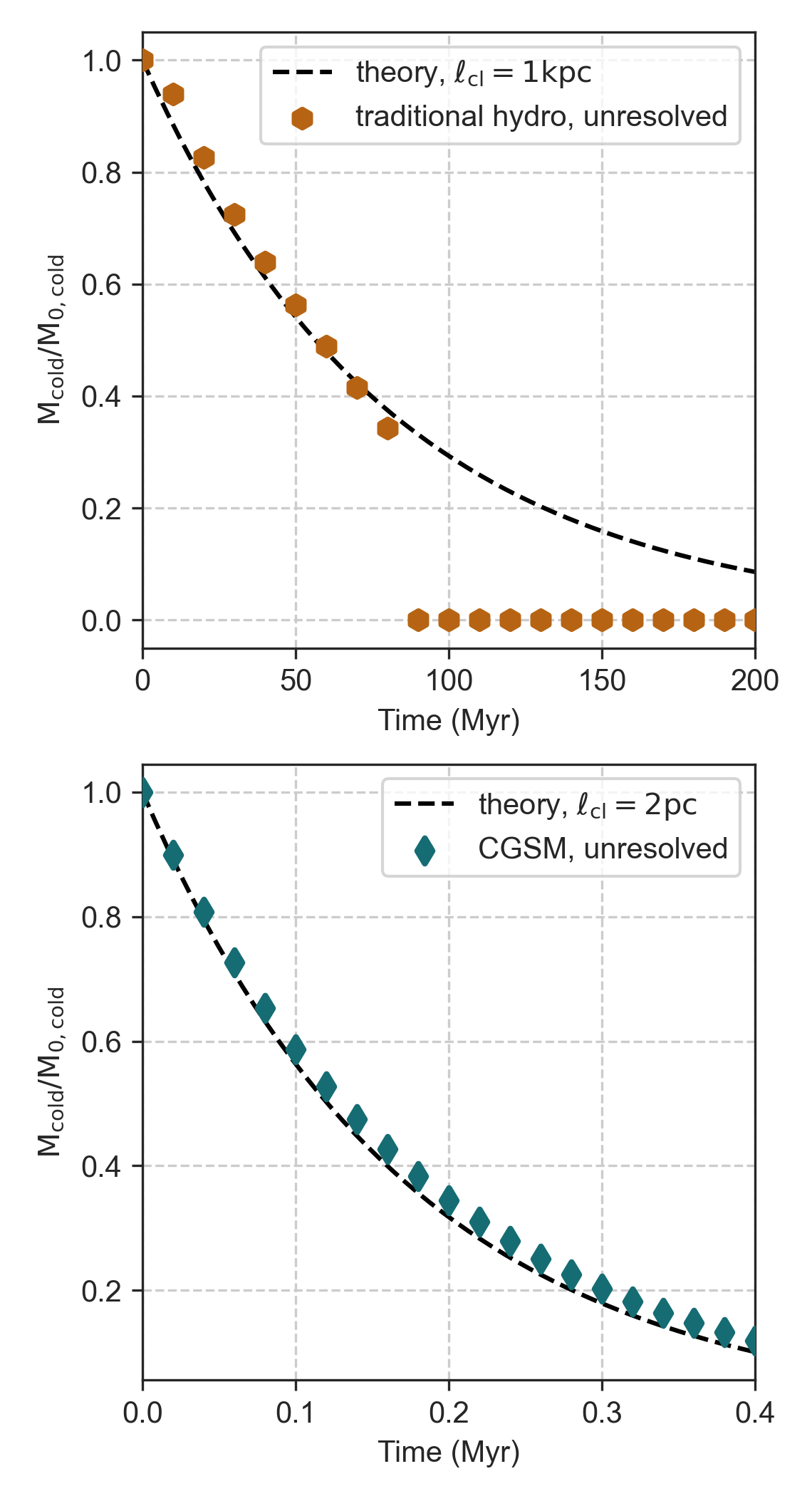}
\caption{The destruction of a cold gas cloud embedded in a hot wind over time in low-resolution simulations using traditional hydrodynamics (top) and the CGSM (bottom). The dashed lines represent the expected mass decay, $M_{\rm cl} / M_{\rm cl, 0} = e^{-t/t_{\rm cc}}$. In the traditional hydrodynamics simulation, the cold cloud's initial size is limited by the resolution to a single cell with $\ell_{\rm cl} = 2\,r_{\rm cl} = 1\, {\rm kpc}$. Conversely, in the CGSM, the cold gas mass is implicitly distributed in unresolved cold cloudlets with $\ell_{\rm cl} = 2\, {\rm pc}$. Although both simulations follow the theoretical curve for at least one cloud-crushing time, the cloud-crushing time in the traditional hydrodynamics simulation is artificially prolonged due to its limited resolution. } 
\label{fig:cloud_crush} 
\end{figure}

\begin{figure}
\includegraphics[width=0.48\textwidth]{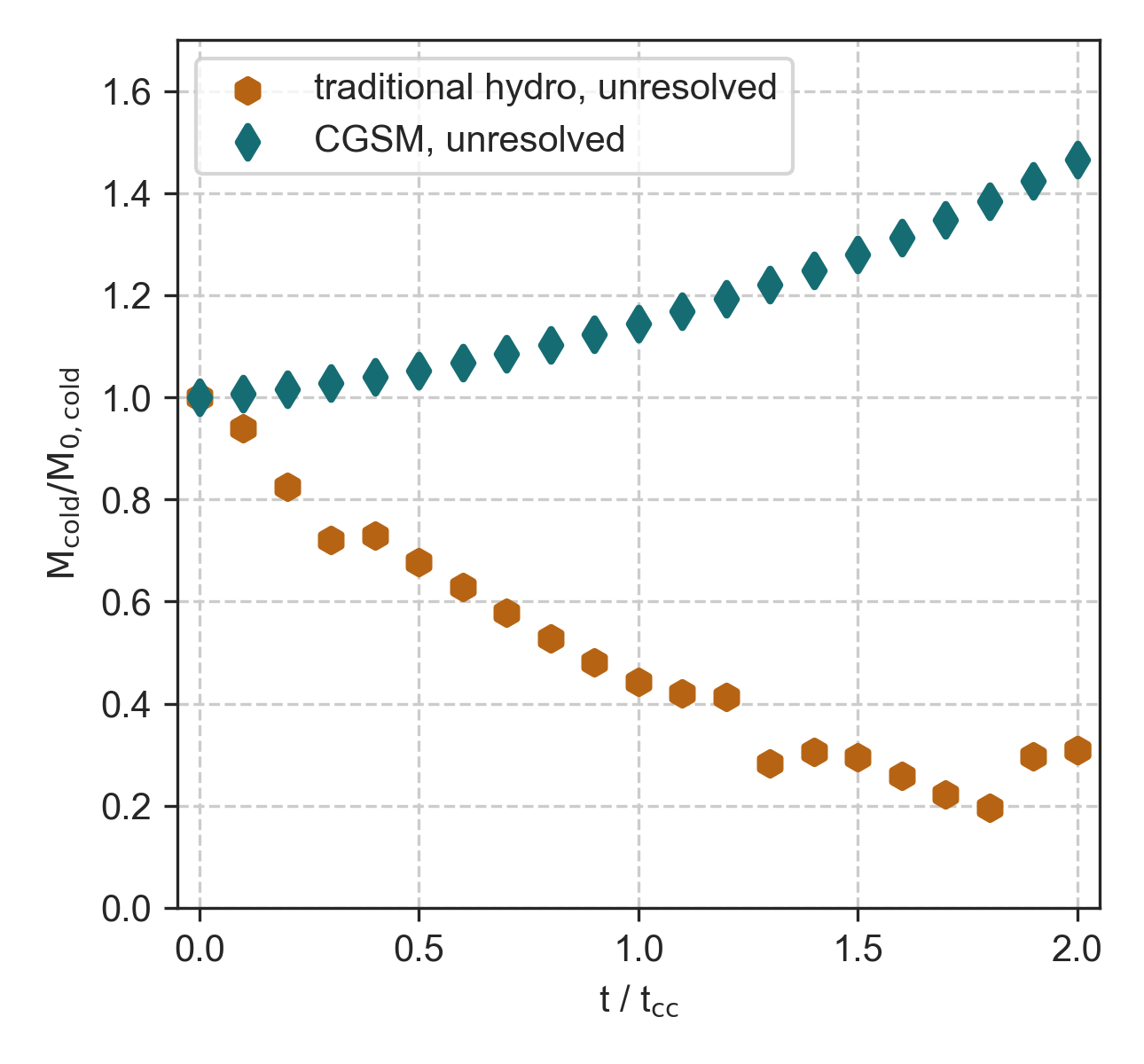}
\caption{The mass evolution of a cold gas cloud embedded in a hot wind in the low-resolution limit. The x-axis shows simulation time normalized by the cloud-crushing time. The initial conditions of these simulations are identical to those in \autoref{fig:cloud_crush}, only with radiative cooling turned on. Since the cooling time of the mixed gas ($T_{\rm mix} = \sqrt{T_gT_{\rm cl}}$) is approximately 30 times shorter than the cloud-crushing time, we would expect the cold gas cloud to accrete mass if it were resolved. Where the traditional hydrodynamics model fails to capture cold cloud mass accretion, the CGSM captures subgrid physics, even at low resolution. 
}\label{fig:cloud_growth} 
\end{figure}

\subsection{Cold Gas Formation in a Single Cell}
For the initial application of our subgrid model, we choose a simple but important scenario: the formation of cold gas in a single cell. The initial conditions consist of a uniform gas with a temperature $T_{g, 0} = 10^6\,{\rm K}$ and density $\rho_{g, 0} = 10^{-27}{\rm g\,cm}^{-3}$. The gas undergoes radiative cooling following an analytic approximation of the cooling curve for a gas with a metallicity of $0.3 Z_{\odot}$. While this metallicity is within the expected range for hot CGM gas \citep{Bregman:2018}, we stress that the exact value has no bearing on the qualitative behavior of the simulations discussed below. The gas is initially motionless, and experiences no external forces. Therefore, when using the CGSM, we focus solely on cloud formation due to thermal instability, omitting terms for cold cloud growth, destruction, and drag.

\autoref{fig:onezone} compares the growth of cold gas mass in both the traditional hydrodynamics (orange) and CGSM (blue) simulations. 
In the traditional hydrodynamic approach, the gas cools uniformly until it the entire cell reaches the temperature threshold ($T < 5.05\times10^4\,{\rm K}$) to be considered as ``cold". As a result, the evolution of cold gas mass follows a step function, centered near the cooling time determined by the initial gas properties. 
In contrast, the CGSM allows for a gradual increase in cold gas mass as it is progressively transferred from the resolved hot phase to the cold fluid. After one cooling cycle, approximately two-thirds of the total gas mass in the CGSM simulation is cold, compared to the traditional model where cold gas constitutes the entirety of the gas mass. After three cooling times, the proportion of cold gas in the CGSM simulation saturates at about 83\% of the total gas mass. 

To provide context, the dashed line shows the cold mass growth rate from an idealized simulation (described in the following section) of thermal instability in which the $\sim 20$ pc cold gas cloudlets are resolved. While there are some quantitative differences between the single-cell and fully-resolved simulations, the latter underscores two key aspects of reality: (1) the growth of cold mass is a gradual process, and (2) the total cold gas mass eventually plateaus, falling short of encompassing 100\% of the total gas mass. Unlike the unresolved hydrodynamic simulation, the CGSM simulation successfully replicates this more realistic behavior.

\subsection{Spatial Distribution of Cold Gas}\label{sec:spatial}
Next, we consider the spatial distribution of cold gas in high-resolution hydrodynamics simulations, low-resolution hydrodynamics simulations, and low-resolution simulations using the CGSM. We simulate thermal instability in idealized 2D patches of CGM-like gas. The simulation setup is described in detail in \citet{Butsky:2020}; however, we summarize the key aspects here. We initialize a $16\times16$ kpc box with a static, uniform gas, setting the initial density at $\rho_{g, 0} = 10^{-27}\,{\rm g\, cm}^{-3}$ and the initial temperature at $T_{g, 0} = 10^6\, {\rm K}$. To seed the thermal instability, we introduce an isobaric perturbation with a white noise spectrum and 2\% amplitude into the uniform medium. In the CGSM, the subgrid cold fluid mass is initialized to zero. We ignore the effects of gravity, focusing solely on the in-situ formation of cold gas.

We model radiative gas cooling with a truncated power law, $\Lambda(T) = \Lambda_0 T^{-2/3}$. This formulation ensures that the temperature of the cold gas remains at $T_{\rm cl} = 5\times 10^4\,{\rm K}$\footnote{Even in idealized, 2D simulations, resolving the full spatial range required for cold gas at $10^4\,{\rm K}$ is computationally expensive. The choice of setting the cold gas temperature to $5\times 10^4\, {\rm K}$ does not qualitatively affect the results.}. We choose the constant, $\Lambda_0 = 1.1\times10^{-19}$ erg cm$^3$ s$^{-1}$ K$^{2/3}$, so that the characteristic radius of cold gas clouds is 10 pc. A crucial aspect of simulating thermal instability involves incorporating a heating mechanism to maintain global equilibrium. We model this heating with a mass-weighted redistribution of the total energy lost to cooling. We refer the interested reader to \citet{Butsky:2020} for additional details.

\autoref{fig:ti_comparison} compares the spatial distribution of cold gas at $t = 2.5\, t_{\rm cool}$ across four different simulations. Progressing from left to right, the first panel shows the fractional cold gas mass ($M_{\rm cl} / M_{\rm cl, total}$) in a traditional hydrodynamics simulation resolved with $4096^2$ cells, corresponding to a spatial resolution of $\Delta x_{\rm cell} = 4\, {\rm pc}$. In this high-resolution simulation, the cold gas appears as a uniformly distributed mist of cloudlets throughout the CGM patch. The second panel depicts the cold gas mass fraction in a traditional hydrodynamics simulation with a much coarser $16^2$ cell resolution, corresponding to a spatial resolution of $\Delta x_{\rm cell} = 1\, {\rm kpc}$, which is typical for the CGM in galaxy-scale simulations. While this lower-resolution simulation generates a total cold gas mass comparable to the high-resolution simulation (with less than a 10\% difference in cold gas masses), there is a marked contrast in the spatial distribution of cold gas clouds. Due to the limited resolution, the minimum size of cold gas clouds is constrained, resulting in fewer but larger clouds compared to the high-resolution simulation. 

The distinct contrast in the spatial distribution of cold gas mass between resolved and underresolved traditional hydrodynamics simulations is further demonstrated in the third panel of \autoref{fig:ti_comparison}. Here, the results of the high-resolution simulation are mapped onto a $16\times16$ grid, mirroring the resolution of the underresolved simulation. This pixelated version of the resolved simulation underscores that, on kiloparsec scales, the distribution of cold gas mass is almost homogeneous, a characteristic that the low-resolution traditional hydrodynamics simulation fails to qualitatively replicate. Conversely, as shown in the right panel, the CGSM is capable of accurately capturing the qualitative distribution of cold gas mass, even at low resolution.

\subsection{Cloud Crushing and Growth}
Next, we turn our attention to the regime in which cold clouds are destroyed within a hot wind, emphasizing how underresolving the CGM in simulations using traditional hydrodynamics can substantially affect the lifespans of cold clouds. 

To illustrate this, we first compare simple cloud-crushing simulations in the unresolved regime with theoretical predictions of cloud-destruction times. The simulation setup is a $64\times16\times16$ kpc box with kpc resolution and periodic boundary conditions. The background is a uniform, hot medium, with $T_g = 10^6\, {\rm K},\, \rho_g = 10^{-28}\,{\rm g\, cm^{-3}}$ and ${\bf v}_g = [100, 0, 0]\,{\rm km\, s^{-1}}$. We place a single cold cloud near the source of the wind. The cold cloud has $T_{\rm cl} = 10^4\, {\rm K},\, \rho_{\rm cl} = 10^{-26}\, {\rm g\, cm}^{-3}$, and is initially at rest. In the context of equation \ref{eqn:cloud_crush}, these initial conditions correspond to $\chi = 100$ and $\xi = 0$, in the limit of no radiative cooling. 

In the traditional hydrodynamics simulation, we replicate a severely underresolved scenario where the total gas mass is concentrated in a single cell, resulting in a cloud of length $\ell_{\rm cl} = 2\,r_{\rm cl} = 1$ kpc. In the CGSM simulation, we distribute the cold gas mass is evenly throughout the simulation domain. Given the temperature of the cold phase and the properties of the hot gas, we expect cold cloudlets to have $r_{\rm cl} = 1\,{\rm pc}$. Both simulations are evolved for two cloud-crushing times.

\autoref{fig:cloud_crush} shows the gradual reduction of cold gas mass over time in both the traditional hydrodynamics simulation (top panel) and the CGSM simulation (bottom panel). At first glance, both models exhibit similar qualitative behavior: the cold gas is progressively destroyed by the hot wind, following the theoretical cloud-crushing timescale (\autoref{eqn:cloud_crush_time}). The low-resolution hydrodynamic simulation abruptly deviates from the analytic solution when the entire cloud is above the threshold temperature ($T_{\rm thresh} = 3\times 10^4\,{\rm K}$) to be considered cold. However the most notable difference in behavior lies in the respective timescales. In the low-resolution traditional hydrodynamics simulation, the destruction time is nearly a thousand times longer due to the artificially enlarged size of the cold cloud. In contrast, the CGSM simulation accurately captures the shorter cloud-destruction timescales, even at the same resolution.

Finally, we consider the low-resolution effects on cloud-growth timescales. To do this, we start with the same initial conditions as in \autoref{fig:cloud_crush} and turn on radiative cooling as described in \autoref{sec:spatial}. In these cloud-crushing simulations, we do not artificially truncate the radiative cooling curve, and allow gas to cool down to $10^4$ K. Given the expected cooling time of the mixed-temperature gas stripped off of the cold cloud ($T_{\rm mix} = \sqrt{T_gT_{\rm cl}}$), these initial conditions correspond to $\xi \approx 30$, placing them firmly in the expected cloud-growth regime \citep{Fielding:2022}. Yet, \autoref{fig:cloud_growth} shows that in the low-resolution limit, the cold cloud in the traditional hydrodynamics simulation loses mass, albeit, more slowly than it did without radiative cooling. While this simulation behavior is expected at low resolutions and small domains \citep{Gronke:2018}, it once again underscores that with $\sim$ kpc resolution, traditional hydrodynamics simulations struggle to capture the correct qualitative behavior of cold CGM gas. Meanwhile, the CGSM allows simulations of the same resolution to capture the subgrid physics of cold mass accretion.

\section{Discussion}\label{sec:discussion}
\subsection{Requirements and Limitations of Simulating Cold Clouds as a Pressureless Fluid}

A key assumption of the CGSM approach is that the unresolved cold clouds are significantly smaller than the size of the resolution element. Therefore, this approach is primarily suited for studying cold gas in the unresolved CGM of galaxy-scale and cosmological simulations. In this context, the implications of modeling the unresolved cold clouds as a pressureless fluid are that we ignore any internal pressure forces in the cold cloud interiors. Also, by imposing pressure equilibrium, we are ignoring any physical processes that introduce pressure anisotropies. For these reasons, the CGSM model is best suited for applications in which the internal pressure anisotropies and dynamics of individual cold clouds is dynamically unimportant.

By modeling cold gas a separate fluid, the CGSM model removes the prohibitively high resolution requirements imposed by resolving the sharp density gradients between the cold and hot gas phases. However, the CGSM still requires that the velocity gradients within each of the fluids are well-resolved. 

\subsection{Comparison to Existing Subgrid Models}
In the last few years, there have been several new subgrid models to address the resolution challenges facing galaxy-scale simulations. Models such as the Physically Evolved Winds \citep[PhEW;][]{Huang:2020, Huang:2022} and Arkenstone \citep{Smith:2024} specifically address the difficulties of accurately simulating multiphase galactic winds driven by supernova feedback. These approaches are motivated by the results of high-resolution, idealized simulations of ISM patches, which have shown that while multiphase outflows predominantly carry their energy in the hot gas phase, the bulk of their mass is in the cold gas phase \citep[e.g.][]{Kim:2020a, Kim:2020b, Li:2020a}. Directly resolving this energy-mass partitioning is currently beyond the capabilities for galaxy simulations in a cosmological context. 

To overcome this limitation, the PhEW and Arkenstone models introduce a novel type of wind particle during supernova feedback events. These wind particles, conceptually similar to the CGSM presented here, represent unresolved cold gas mass and engage in mass, momentum, and energy exchanges with the surrounding medium before they are ultimately recoupled with the ``regular'' particles in the CGM. 

A key difference between the PhEW and Arkenstone models and the CGSM lies in the spatial extent and continuity of the cold gas subgrid prescription. In the PhEW and Arkenstone frameworks, cold wind particles are generated exclusively during supernova feedback events. Once these particles recouple with the regular gas, the capacity to model subgrid cold gas ceases. In contrast, the CGSM allows every cell in the simulation to host arbitrarily small amounts of cold gas. This approach facilitates evolving a continuous distribution of subgrid cold gas throughout the entire simulation domain. Furthermore, the grid-based approach to the CGSM is particularly well-suited for solving the conservation equations in low-density regions.

\subsection{Missing Physics and Future Applications}
In this work, the CGSM intentionally omits certain physical processes, such as magnetic fields, conduction, turbulence, and cosmic rays, to focus on presenting a ``proof-of-concept'' of the advantages of a subgrid approach in CGM simulations.
While this missing physics would likely alter the quantitative aspects of the cold-hot gas interactions -- such as the rate of thermal instability growth or the timescales for cloud mass loss and accretion -- the modular design of the CGSM makes it straightforward to incorporate changes to the expressions for $\dot{\rho},\, \dot{\bf p}$, and $\dot{\varepsilon}$. Future modifications will incorporate new physics or reflect updated methodologies based on the latest high-resolution simulations, such as using a power-law to model unresolved cold cloud sizes or accounting for the turbulent velocity dispersion of unresolved clouds within a cell. 

Importantly, changes to the quantitative details of the interaction terms do not compromise the qualitative advantages offered by the CGSM over standard hydrodynamics simulations in the low-resolution limit. The CGSM's ability to (1) capture the gradual accretion of cold gas mass within a single cell, (2) generate smooth spatial distributions of cold gas, and (3) accommodate short cloud-destruction and growth timescales remain fundamentally robust, irrespective of these potential updates. 

The assumption in the CGSM that subgrid cold gas clouds are significantly smaller than the typical CGM resolution ($\Delta x \sim 1$ kpc) may not hold scenarios where extreme nonthermal pressure leads to the formation of significantly larger cold ``clouds'' with characteristic sizes larger than $\sim 1-100$ kpc, as seen in some CGM simulations that include cosmic-ray physics \citep[e.g.][]{Salem:2016, Butsky:2018, Buck:2020, Ji:2020, Butsky:2022}. However, in such cases, the nonthermal pressure can be factored into the approximation of the expected cold cloud sizes, as demonstrated in \citealt{Butsky:2020}. When cold clouds are large enough to be resolved at a given resolution, the CGSM does not need to be applied in that region, and the mass of the cold subgrid fluid would simply be zero.

In future work, we plan to incorporate the missing physics described above and calibrate the CGSM for use in cosmological zoom-in simulations. The combination of a cosmological context and physically motivated treatment of subgrid cold-gas physics will enable us to better determine the origin and impact of cold gas in a variety of contexts, including the CGM, galactic winds and mass accretion, as well as high-velocity clouds in our own Milky Way.

\section{Summary}\label{sec:summary}
In response to the challenge of resolving cold CGM gas in galaxy simulations, we introduce a two-fluid framework for modeling the subgrid physics of unresolved cold gas. The CGSM is designed to explicitly evolve the total mass density and bulk momentum of unresolved cold gas cloudlets. It uses the properties of the resolved gas fluid to inform predictions about the physical state of cold gas. In this model, the unresolved cold fluid interacts with the resolved hot fluid, exchanging mass, momentum, and energy in accordance with the findings from high-resolution, idealized simulations (\autoref{fig:diagram}).

The CGSM offers several distinct benefits over traditional hydrodynamics methods in situations where the resolution is significantly lower than necessary to adequately resolve cold-gas structure. In contrast to traditional hydrodynamics simulations, which are limited to a single-phase, single-temperature gas within each cell, the CGSM allows for the presence of arbitrarily small amounts of cold gas throughout the simulation (\autoref{fig:onezone}). As a result, the CGSM is capable of producing more realistic spatial distributions of cold gas mass. This is in stark contrast to underresolved traditional hydrodynamics simulations, which tend to accumulate cold gas mass in a limited number of large clouds, with sizes artificially inflated by the size of the low-resolution voxels (\autoref{fig:ti_comparison}). Furthermore, where underresolved hydrodynamics simulations predict artificially long cloud-destruction and accretion timescales, the CGSM captures the expected behavior of cold gas, even when operating at the same resolution (\autoref{fig:cloud_crush}, \autoref{fig:cloud_growth}). 

These findings suggest that in the limit of low resolution -- as is typical in the halos of galaxy-scale simulations -- traditional, single-fluid hydrodynamic simulations may be unreliable tools for determining the origin and evolution of cold CGM gas. Even if the simulations converge on certain cold-gas metrics, such as the total cold gas mass, by artificially inflating cold-gas sizes and evolution timescales, we cannot rule out that such simulations are finding the ``right answer'' for the wrong reasons. 

Certainly, opting for a subgrid model comes with its own set of trade-offs. Fundamentally, this approach introduces new simulation parameters that require precise tuning, nuanced resolution requirements, and the assumption that the subgrid model accurately represents the ensemble of cold clouds. Using a subgrid model for cold CGM gas also means that the simulations cannot be used to study small-scale cold gas structure and evolution. However, for simulations with a prohibitively large dynamic range of physical and temporal scales, the subgrid approach is unavoidable. For example, there is a strong precedent for using subgrid models of star formation and stellar feedback in galaxy-scale simulations. While such simulations cannot be used to study stellar evolution or supernova remnants, the subgrid approach has been invaluable for understanding the effects of star formation and stellar feedback on galaxy evolution. We are now faced with a similar trade-off in studying galactic halos. 

For those seeking to study the flow of cool gas and its relationship to galaxy evolution on cosmological scales, the subgrid model approach is likely inevitable. The alternative -- inferring cold gas properties from simulations where the resolution elements are orders-of-magnitude larger than the actual cold gas structures -- is fundamentally flawed. Choosing the right approach hinges on a better understanding of which scales need to be resolved in order to accurately model cold-gas physics. Should it be determined that the required scales are small ($\lesssim$ pc) compared to the typical resolution in the simulated CGM ($\gtrsim$ 100s pc), then directly resolving cold CGM gas would be computationally infeasible with current technologies. In anticipation of such a scenario and rather than waiting potentially decades for the requisite computational advancements, the CGSM offers a means to effectively model cold-gas physics within the limitations of current resolution capabilities.

\acknowledgments{We thank the referee, Rainer Weinberger, for constructive comments and suggestions that improved this manuscript. The authors would also like to thank Greg Bryan, Drummond Fielding, Matthew Smith, and Peng Oh for insightful conversations that contributed to the development of ideas presented in this paper. I.S.B. was supported by HST Legacy grant AR-15800, the DuBridge Postdoctoral Fellowship at Caltech, and by NASA through the Hubble Fellowship, grant HST-HF2-51525.001-A awarded by the Space Telescope Science Institute, which is operated by the Association of Universities for Research in Astronomy, Incorporated, under NASA contract NAS 5-26555. C.B.H. is supported by NSF grant AAG-1911233, and
NASA grants 80NSSC23K1515, HST-AR-15800, HSTAR-16633, and HST-GO-16703. Support for P.F.H. was provided by NSF Research Grants 1911233, 20009234, NSF CAREER grant 1455342, NASA grants 80NSSC18K0562, HST-AR-15800.001-A. }

\section*{software}
The CGSM was implemented in the {\sc enzo} astrophysical simulation code \citep{Bryan:2014, Enzo:2019}. The analysis of the simulations relied heavily on the {\sc yt} \citep{Turk:2011}, {\sc matplotlib} \citep{matplotlib}, and {\sc numpy} \citep{numpy} packages for the {\sc python} \citep{ipython} programming language. 

\datastatement{The data supporting this article and CGSM source code are available on reasonable request to the corresponding author.} 

\bibliography{main}

\begin{figure*}
\includegraphics[width=\textwidth]{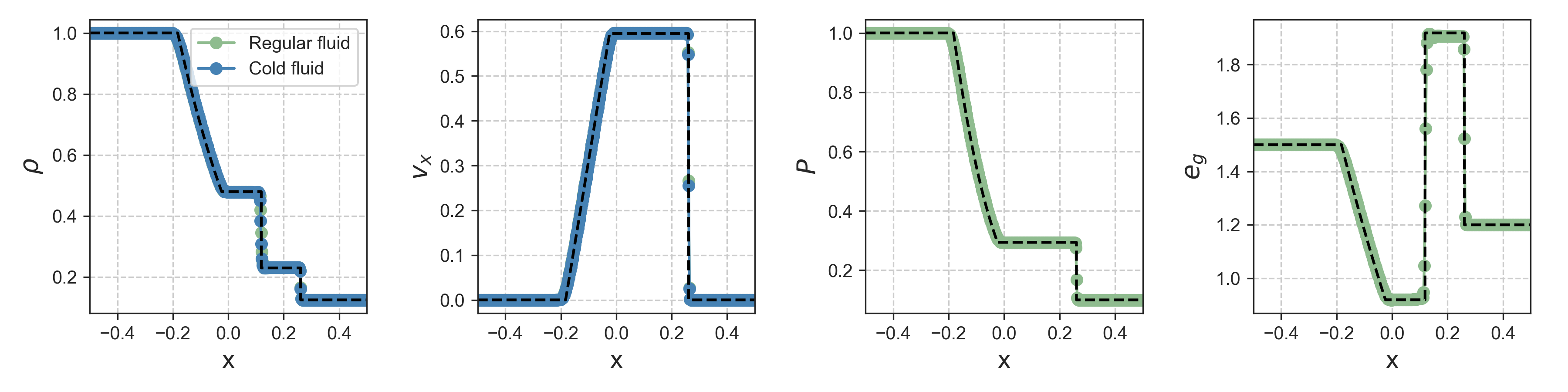}
\includegraphics[width=\textwidth]{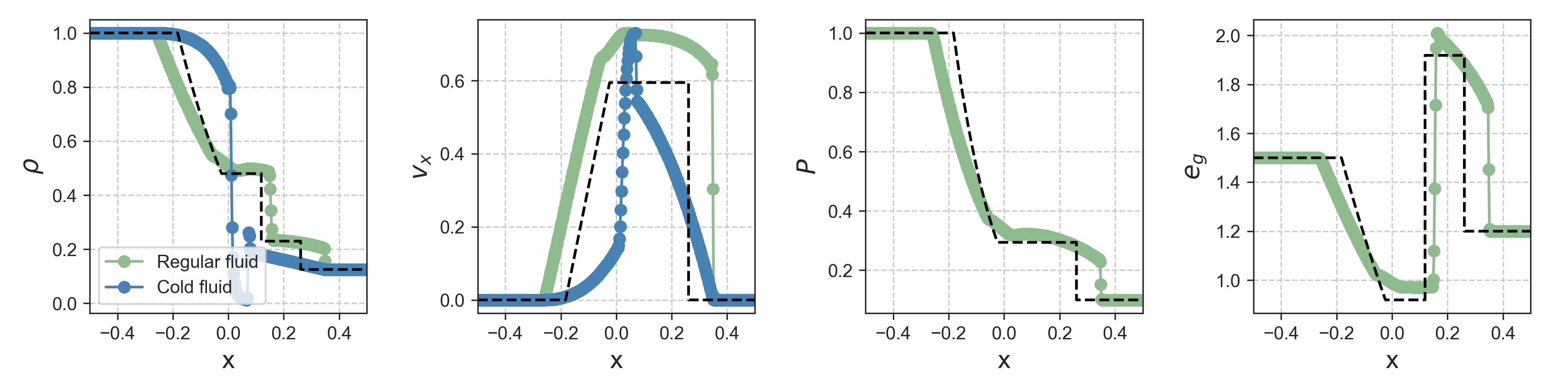}
\caption{ The distribution of the gas density, velocity, pressure, and specific thermal energy in a one-dimensional modified SOD shocktube after $t = 0.2$ code time units, for two different drag coefficients. {\bf Top:} When the drag coefficient is very high ($K_{\rm drag}$ = 1000), the cold and regular gas fluids move at the same velocity, following the analytic solution for a shocktube with a modified sound speed (black dashed line). {\bf Bottom:} When the drag coefficient is only moderately strong ($K_{\rm drag}$ = 1), the shock motion of the hot fluid imparts momentum to the cold gas fluid, but the two fluids are not fully coupled. Although there is no analytic solution for this regime, our results are consistent with those presented in \citet{Laibe:2012a}.} 
\label{fig:shocktube} 
\end{figure*}

\appendix

\section{Tests of Model Behavior}\label{sec:appendix}
In this section, we test the core functionality of the two-fluid model and its implementation in the {\sc enzo} astrophysical simulation code \citep{Bryan:2014, Enzo:2019}.

In \autoref{fig:shocktube}, we demonstrate the advection of the two-fluid model in the presence of a shock. For this test, we set up a modified Sod shocktube with both the regular fluid and the unresolved cold fluid initialized to the same values described below. We simulate the shocktube in a 1D domain with $x \in [-0.5, 0.5]$, resolved by 200 cells. The initial conditions are given by $\rho_g = 1, \bar{\rho}_{\rm cl} = 1, P_g = 1$ for $x \le 0$, and $\rho_g = 0.125, \bar{\rho}_{\rm cl} = 0.125, P_g = 0.1$ for $x \ge 0$. The velocity of both the regular fluid and the cold subgrid fluid is initialized to zero everywhere. $P_g = (\gamma - 1)\varepsilon$ is the thermal gas pressure with $\gamma = 5/3$. We evolve the shocktube for $t = 0.2$ internal time units. 

In the case of strong drag ($K_{\rm drag} = 1000$), the analytic solution is given by the black dashed line. In this case, the simulated shocktube follows the analytic solution well. There is no analytic solution for the case of weak drag. Instead, we repeat the numerical test in \citet{Laibe:2012a} with $K_{\rm drag} = 1$ and plot the results in the bottom panel of Figure \ref{fig:shocktube}. The behavior of the two fluids agrees well with the results in \citet{Paardekooper:2006} and \citet{Laibe:2012a}.

\begin{figure}
\includegraphics[width=0.5\textwidth]{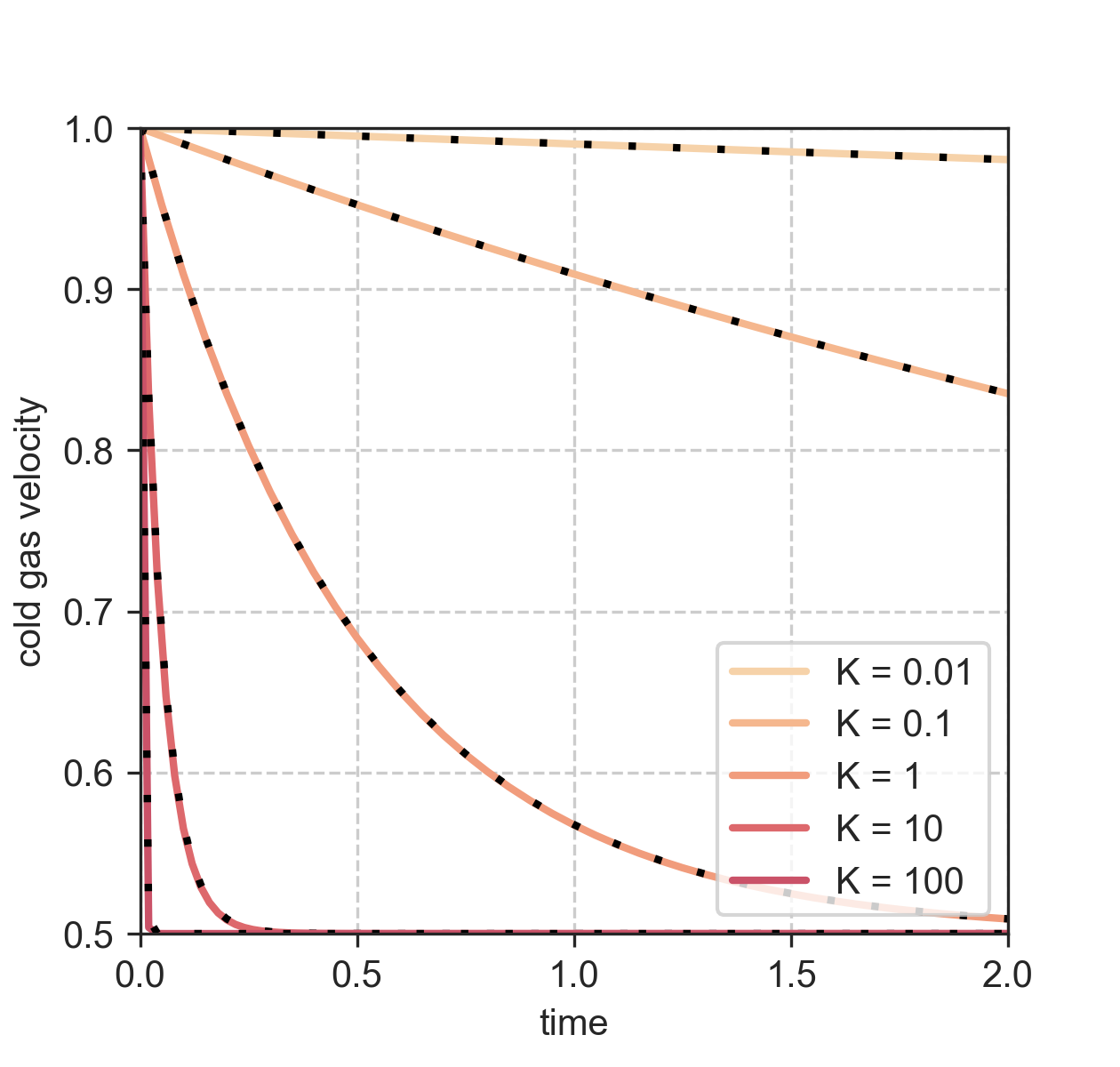}\\
\includegraphics[width=0.5\textwidth]{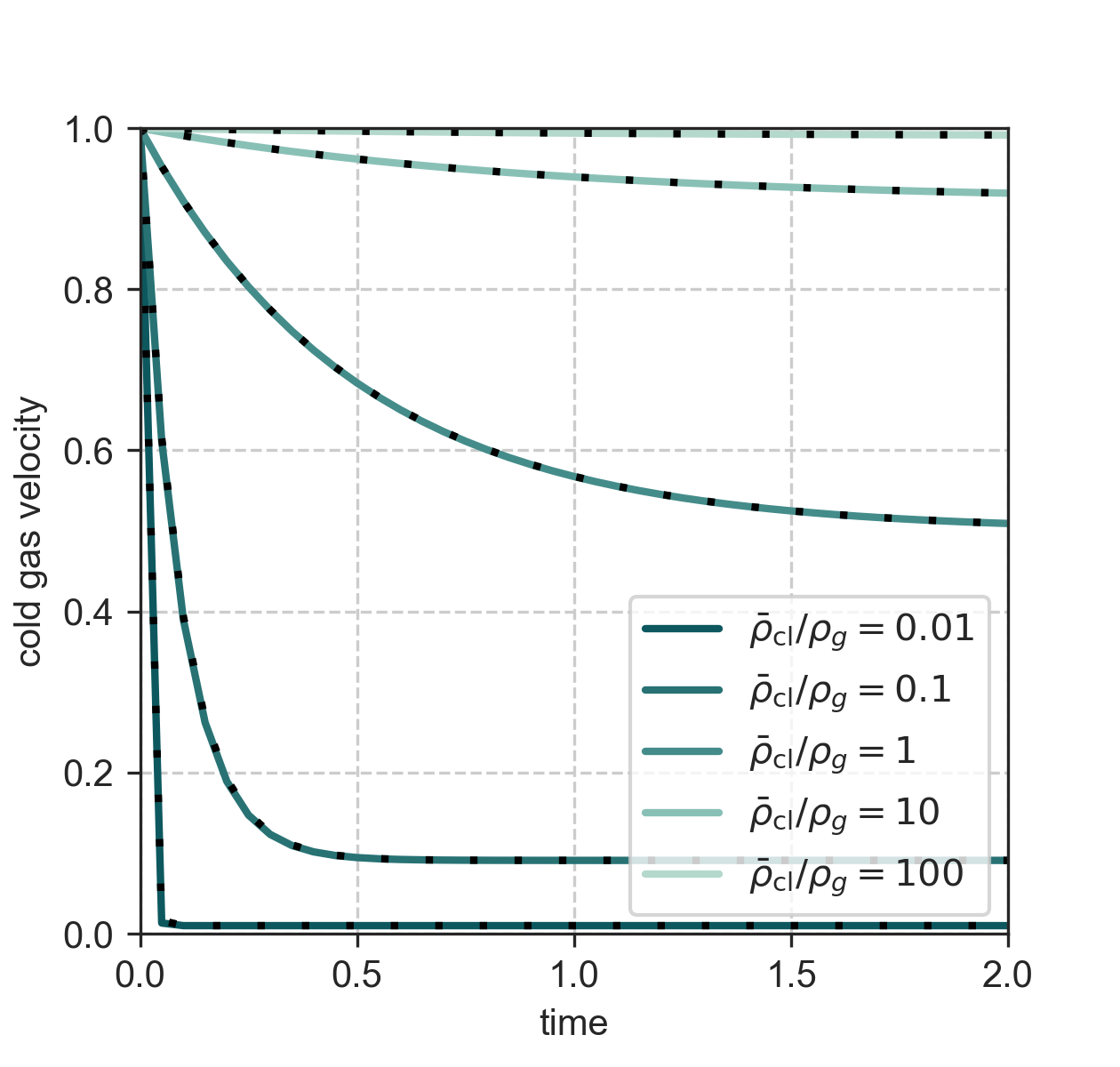}
\caption{ The time evolution of cold gas velocity for a variety of different drag coefficients (top) and ratios of cold gas density to regular gas density (bottom). The black dotted lines show the analytic solutions, which agree with the simulation results in all cases. When the drag coefficient is low or the cold gas density is high relative to the hot gas density, the velocity of the cold gas remains relatively unchanged over time. When the drag coefficient is high or the cold gas density is low, the cold gas velocity quickly reaches its equilibrium velocity. }
\label{fig:dragtest} 
\end{figure}
 
In \autoref{fig:dragtest}, we demonstrate the performance of the drag coupling with a modified version of the \emph{dustybox} test in \citet{Laibe:2012a}. The physical setup of this problem is similar to that of the shocktube, only the gas properties are uniform throughout the entire space and the boundary conditions are periodic. The normal gas is initially at rest and the cold fluid is initialized with a velocity in the $\hat{x}$ direction. If the cold and regular fluids are coupled through a drag coefficient, then the velocity of the cold gas will decrease over time as it imparts momentum on the regular gas. 
For this test, we consider the simplest case, in which the drag coefficient, $K_{\rm drag}$, is a constant.

We simulate this process for a variety of different drag coefficients, $K_{\rm drag}$, and cold gas density ratios, $\bar{\rho}_{\rm cl} / \rho_{\rm g}$. In all cases, the initial hot gas density and the velocities of the two fluids are: $\rho_{\rm g} = 1, {\bf v}_{g} = 0, {\bf \bar{v}_{\rm cl}} = 1$. When testing the effect of the drag coefficient, we keep $\bar{\rho}_{\rm cl} / \rho_{\rm g} = 1$ constant and vary $K_{\rm drag} \in [0.01, 0.1, 1, 10, 100]$. When testing the effect of the cold gas density ratio, we fix $K_{\rm drag} = 1$ and vary $\bar{\rho}_{\rm cl} / \rho_{\rm g} \in [0.01, 0.1, 1, 10, 100]$. 

\autoref{fig:dragtest} shows the time evolution of the cold gas velocities for a variety of different drag coefficients (top) and ratios of cold gas density to regular gas density (bottom). The black dotted lines show the analytic solutions, which agree with the simulation results in all cases. When the drag coefficient is low or the cold gas density is high relative to the hot gas density, the velocity of the cold gas remains relatively unchanged over time. When the drag coefficient is high or the cold gas density is low, the cold gas velocity quickly reaches its equilibrium velocity.

% Don't change these lines
%\bsp	% typesetting comment
\label{lastpage}
\end{document}